\documentclass[aps,
twocolumn,
]{revtex4} 
\usepackage{epsf}
\usepackage{amssymb,amsmath}
\usepackage{hyperref}
\usepackage{graphicx}

\begin{document}
\title{Adsorption of Finite Polymers in Different Thermodynamic Ensembles}
\title{Systematic Microcanonical Analyses of Polymer Adsorption Transitions}
\author{Monika M\"oddel}
\email[E-mail: ]{Monika.Moeddel@itp.uni-leipzig.de}
\author{Wolfhard Janke}
\email[E-mail: ]{Wolfhard.Janke@itp.uni-leipzig.de}
\homepage[\\ Homepage: ]{http://www.physik.uni-leipzig.de/CQT.html}
\affiliation{Institut f\"ur Theoretische Physik,
Universit\"at Leipzig, Postfach 100\,920, D-04009 Leipzig,\\
and Centre for Theoretical Sciences (NTZ), Emil-Fuchs-Stra{\ss}e 1, D-04105 Leipzig, Germany}
\author{Michael Bachmann}
\email[E-mail: ]{bachmann@smsyslab.org}
\homepage[\\ Homepage: ]{http://www.smsyslab.org}
\affiliation{Institut f\"ur Festk\"orperforschung, Theorie II,\\
Forschungszentrum J\"ulich, D-52425 J\"ulich, Germany}%
\begin{abstract}
We investigate the cooperative effects of a single finite chain of monomers near an attractive substrate
by first constructing a conformational pseudo-phase diagram based on the
thermal fluctuations of energetic and structural quantities. Then, the
adsorption transition is analyzed in more detail. This is conveniently done
by a microcanonical analysis of densities of states obtained by extensive
multicanonical Monte Carlo simulations. 
For short chains and strong surface attraction, the microcanonical entropy 
turns out to be a convex function of energy in the transition regime.
This is a characteristic physical effect and deserves a careful consideration
in analyses of cooperative macrostate transitions in finite systems.
\end{abstract}
\maketitle

\section{Introduction}
  The understanding of the adsorption phenomena of polymers on surfaces is a prerequisite, e.g., 
  for designing micro- or nanostructures. 
  Also the fact that various polymers are usually found near both sides of cell membranes and are important for their mechanical
  stability and physiological function~\cite{membrane} has driven studies of polymers near surfaces and
  interfaces~\cite{membrane1}. 
  In this context a deeper understanding of the origin of specific binding affinities
  of proteins regarding the type of the substrate and the amino acid sequence is very desirable. 
  In recent years, some progress has been made in this field~\cite{sarikaya,goede}, but due to the complexity 
  introduced by the huge amount of possible sequences and surfaces many problems are still open. 
  A qualitative understanding of the cooperative nature of the adsorption transition for short chains can, however,
  already be gained by studying the behavior of homopolymers close to a flat substrate~\cite{eisenriegler1,prellberg,binder}.
  We focus here on the systematic description of the phase diagram in a wide parameter range and cooperative effects
  of chains of finite length.
  
  First, in this numerical study, the conformational pseudo-phase diagram of a coarse-grained non-grafted off-lattice
  polymer will be constructed versus temperature 
  and surface attraction strength. The competition between monomer-monomer and surface-monomer
  attraction gives rise to a variety of different conformational phases~\cite{michael,mbj1}. Our computer simulations rely on the multicanonical Monte Carlo 
  method~\cite{muca1} that allows for the precise determination of the canonical expectation values of suitable 
  observables over a wide range of temperatures within a single long simulation run. 
  In addition it yields an estimate of the density of states,
  which possesses a convex regime at the adsorption transition in the case of short chains and strong surface attraction. 
  Albeit known to be a continuous transition in the thermodynamic limit of infinitely long chains\cite{eisenriegler1}, the adsorption transition of 
  non-grafted finite-length polymers thus exhibits a clear signature of a first-order-like transition, with coexisting
  phases of adsorbed and desorbed conformations.

\section{Model and Simulation}
\textit{Off-Lattice Homopolymer with Attractive Substrate.}
  We employ a coarse-grained off-lattice model for 
  homopolymers that has also been generalized for studies of
  heteropolymers~\cite{stillinger1} and helped to understand protein folding
  channels from a mesoscopic perspective~\cite{stefan1}. 
  Adjacent monomers are connected by rigid bonds
  of unity length, but bond and torsional angles are free to
  rotate. The energy function consists of three terms,
  \begin{eqnarray}
  E & = &\, 4\sum_{i=1}^{N-2}\sum_{j=i+2}^{N}\left( {r_{ij}^{-12}}-{r_{ij}^{-6}}\right)+
  \frac{1}{4}\sum_{i=1}^{N-2}\left[ 1-\cos\left(\vartheta_{i} \right)\right]\nonumber\\ 
   & &+
  \epsilon_{s}\sum_{i=1}^{N}\left(\frac{2}{15} {z_i^{-9}} - {z_i^{-3}}\right),
  \end{eqnarray}
  where the first two terms give the energy of a polymer in bulk ($E_{\rm bulk}$) 
  that consists of the standard 12-6 Lennard-Jones (LJ) potential and a weak bending energy.
  The bending energy provides a penalty for successive bonds deviating from a straight arrangement.
  Here $0\leq \vartheta_{i}\leq\pi$ denotes the bending angle between monomers $i$, $i+1$, and $i+2$.
  The distance between the monomers $i$ and $j$ is 
  ${r}_{ij}$ and $z_i$ is the distance of the $i$th monomer to the substrate.
  The third term is the attractive surface potential $E_{\rm surf}$, obtained by integrating over 
  the continuous half-space $z<0$, where every space ele\-ment interacts with 
  a single monomer by the usual 12-6 LJ expression\cite{steele1}. 
  Hence, the parameter $\epsilon_s$ weighs the 
  monomer-surface ($E_{\rm surf}$) and monomer-monomer ($E_{\rm bulk}$) interaction. 
  Center-of-mass translation is restricted by the attractive substrate
  at $z=0$ and a sufficiently distant steric wall at $z=L_z$.
  In our microcanonical analysis it will become clear how $L_z$ influences the results,
  however, the effect on the canonical data is small if $L_z$ exceeds the extension of the polymer.
  As long as not mentioned otherwise, the ratio $N/L_z$ is kept constant ($L_z=3N$).  
  We always employ natural units ($k_B\equiv 1$).
  
\textit{Energetic and Structural Quantities.}
  To describe the canonical equilibrium behavior, we use the 
  canonical expectation values and thermal fluctuations of the 
  following quantities:
  energy and  specific heat, $c_{V}$, 
  the radius of gyration, $\left<R_{\rm gyr}\right >$, as a measure for the extension of the polymer,
  and its tensor components parallel and perpendicular to the surface, $\left<R_{\parallel}\right >$
  and $\left<R_{\perp}\right >$, with $R_{\rm gyr}^{2}=R_{\parallel}^{2}+R_{\perp}^{2}$. 
  The components are of interest due to the structural anisotropy introduced by the substrate.
  Other useful quantities are the 
  distance of the center-of-mass of the polymer to the surface, $\left<z_{\rm cm}\right >$,
  and the mean number of monomers docked to the surface. 

\textit{Multicanonical Sampling.}
  The density of states $g(E)$ encodes all information regarding the phase behavior
  of the system rendering its precise estimation extremely helpful. 
  This requires the application of sophisticated
  Monte Carlo methods.
  In this work, we have performed multicanonical simulations~\cite{muca1}.
  The idea is to increase the sampling
  rate of conformations being little favored in the free-energy landscape 
  by performing a random walk in energy space. This is achieved 
  by introducing suitable multicanonical weights $W_{\rm muca}(E)\sim g^{-1}(E)$
  to sample conformations $\bf X$ according to a transition probability
  \begin{equation}
  \omega(\mathbf{X}\rightarrow \mathbf{X'})=\min [W_{\rm muca}(E(\mathbf{X'}))/W_{\rm muca}(E(\mathbf{X})),1].
  \label{eq:transition}
  \end{equation} 
  As the weights $W_{\rm muca}(E)$, i.e.~$g(E)$, are unknown a priori, they are determined
  iteratively until the energy histogram is constant up to a variation of about $10\%$ 
  in the desired energy range. An efficient, error-weighted multicanonical recursion 
  is described in Ref.~\cite{muca2}.

\section{Pseudo-Phase Diagram}\label{pseudodiscussion}
\begin{figure}[t!]
  \begin{center}	\includegraphics[width=8.3cm]{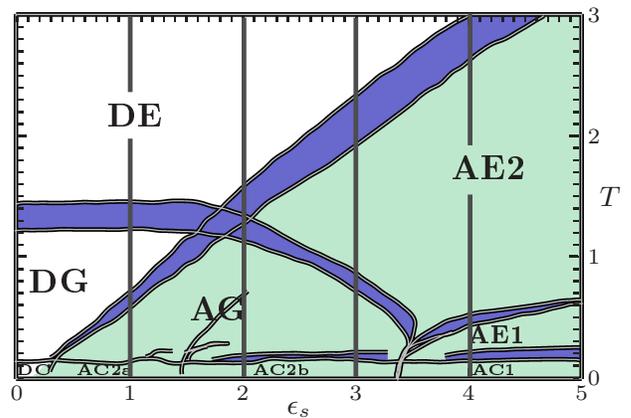}\end{center}
  \caption{\label{fig:diagram}Pseudo-phase diagram of the 20mer (for details, see text). Different representative conformations are shown in Fig.~\ref{fig:5}.
  Along the lines of constant $\epsilon_s$ a 
  microcanonical analysis has been performed, see section \ref{microsection}.}
  \end{figure}

\begin{figure}[h]
  \begin{center}	\includegraphics[width=7.0cm]{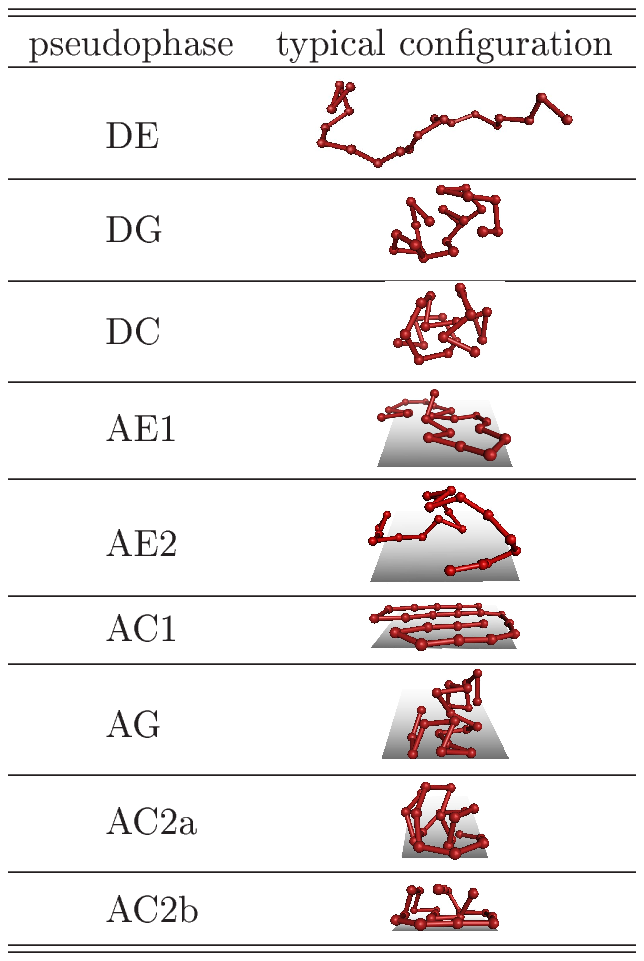}\end{center}
	  \caption{\label{fig:5} Representative examples of conformations for the 20mer in
  the different regions of the $T$-$\epsilon_s$ pseudo-phase diagram in Fig.~\ref{fig:diagram}. DE, DG, and 
  DC represent desorbed ``phases''.
  In regions AE1, AE2, AC1, AG, AC2a, and AC2b, conformations are favorably adsorbed.}
  \end{figure}

  To construct the conformational pseudo-phase diagram, multicanonical
  simulations~\cite{muca1} for 51 different surface attraction strengths 
  $\epsilon_s\in [0,5]$ were performed for a chain with $N=20$ monomers. These data can now be reweighted to arbitrary temperature,
  but since it turns out that the interval $T\in(0,3]$ is the most interesting one, we restrict ourselves to this range here. 
  Each simulation consisted of $10^8$ sweeps and was 
  performed with at least two different initializations.
  The final pseudo-phase diagram is shown in Fig.~\ref{fig:diagram} and representative conformations are given in Fig.~\ref{fig:5}. 
  The blue bands indicate the approximate phase boundaries that have some uncertainty
  because the peaks of the fluctuation of canonical expectation values do not coincide for
  finite systems. 
  It should be stressed that due to the finite chain lengths all phases and transitions 
  here are not phases in the strict thermodynamic sense.
  Nevertheless, a reasonable picture of polymer adsorption behavior is obtained and most of the 
  phases are believed to still exist for longer chains.
  Here, only some representative observables used for the construction will be discussed. For more details, see
  Ref.~\cite{mbj1}.

\textit{Energetic Fluctuations.}
  Although the energy varies smoothly with $T$ and $\epsilon_s$, 
  two transitions can be identified as ridges in the profile of 
  the specific heat: The \textit{adsorption transition} separating desorbed and 
  adsorbed conformations and a \textit{freezing transition} 
  at low temperatures. Near $T=0.25$, $c_V$ exhibits a pronounced peak 
  independently of $\epsilon_s$.
  The crystalline shape of the structures below this peak additionally
  confirms its nature as freezing transition.
  However, to identify different crystalline shapes, a closer look at the 
  conformational quantities is needed.

  \begin{figure}
		  \begin{center}\includegraphics[width=8.7cm]{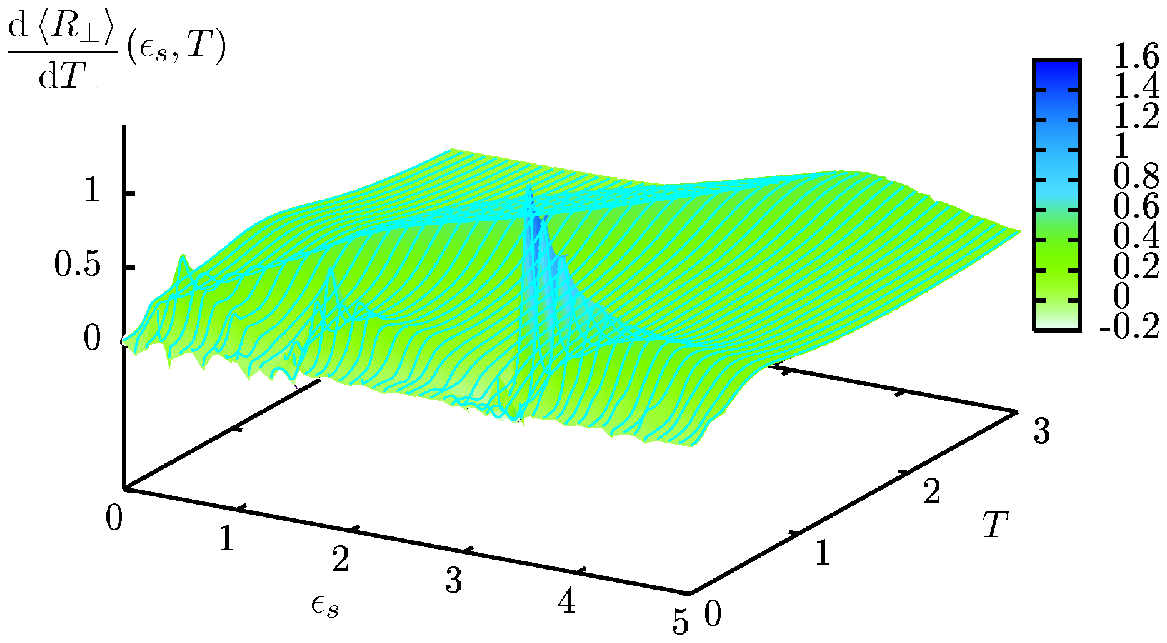} 		\end{center}
  \caption{\label{fig:RgyrzE}  ${\rm d}\left<R_{\perp}\right>/{\rm d}T$ of the 20mer.  }
  \end{figure}

\textit{Structural Fluctuations.}
  The average radius of gyration $\left\langle R_{\rm gyr} \right\rangle$  reveals that the most compact conformations dominate at low 
  $T$ and low $\epsilon_s$. It establishes the phase boundaries between DE (desorbed expanded) and DG (desorbed globular) and between
  AE2 (adsorbed extended; not flat on substrate) and AG (adsorbed globular) and confirms the freezing transition, but the 
  adsorption transition is 
  not prominently signaled by $\left\langle R_{\rm gyr} \right\rangle$.
  Its tensor components give additional information.
  For example, for $\epsilon_s \geq3.4$, $\left\langle R_{\perp} \right\rangle $ vanishes at low 
  $T$, whereas $\left\langle R_{\parallel} \right\rangle $ attains low values at 
  lower $\epsilon_s$. Small values of $\left\langle R_{\perp} \right\rangle $ correspond to conformations
  spread out flat on the surface, with associated pseudo-phases
  AC1 (adsorbed compact; flat) and AE1 (adsorbed expanded; flat),
  separated by the freezing transition. 
  The most pronounced transition is the strong layering 
  transition at $\epsilon_s\approx 3.4$ that separates regions of planar 
  conformations (AC1, AE1) from the region of 
  double-layer structures (AC2b) and adsorbed globules 
  (AG), below and above the freezing transition, respectively. 
  This sharp energe\-tical transition
  can, e.g., be nicely seen in ${\rm d}\left<R_{\perp}\right>/{\rm d}T$ in Fig.~\ref{fig:RgyrzE}.
  Although for the considered short chains no higher-layer
  structures are observed, 
  $\left\langle R_{\parallel,\perp} \right\rangle $ indicate some 
  activity for lower $\epsilon_s$.
  For $N=20$, $\epsilon_s\approx1.4$ is the lowest attraction strength, 
  where stable double-layer conformations are found.
  What follows is a low-temperature subphase of surface attached 
  compact conformations (AC2a). 
  These structures occur if the surface attraction is not strong enough 
  to induce the formation of compact layers. 
  Structures here are subject to quite strong finite-size effects.
  Raising the temperature above the freezing transition starting in the AC2 regions, 
  polymers adopt the adsorbed, globular, but unstructured conformations
  of the AG phase.
  This pseudo-phase has been first conjectured from short exact enumeration studies 
  of 2D polymers in poor solvent~\cite{kumar1}, but was also found in lattice-polymer simulation
  studies\cite{prellberg,michael}. 
  At even higher $T$, two scenarios
  can be distinguished depending on the relative strengths of
  $E_{\rm bulk}$ and $E_{\rm surf}$. For low $\epsilon_s$, the polymer first desorbs 
  (from AG to DG) and expands
  at even higher tempera\-tures (from DG to DE). For larger $\epsilon_s$, the polymer expands while 
  it is still adsorbed (from AG to AE2) and desorbs at higher 
  $T$ (from AE2 to DE). 
  The remaining observables confirm the picture sketched so far. 
  The center-of-mass distance to the surface $\left\langle z_{\rm cm}\right\rangle $ 
  gives a clear signal of the adsorption transition, whose location is well described by
  $T_{\rm ads} \propto \epsilon_s$. Since at 
  higher $T$ the stronger thermal fluctuations are more likely to 
  overcome the surface attraction, this is intuitive.
  The mean number of surface contacts supports the observed layering.

  It is clear that in particular in the
  compact pseudo-phases the structural behavior of the studied small
  chains is affected by finite-size effects.
  However, especially at high temperatures, the pseudo-phase diagram constructed here corresponds quite well with a similar 
  lattice study\cite{michael} with the advantage of not suffering from lattice artifacts. 

\section{The Adsorption Transition Revisited Microcanonically}\label{microsection}
  We now concentrate on the adsorption transition and look at it from another perspective:
  the microcanonical one. This approach has already proven quite useful for 
  first-order-like structural transitions such as
  molecular aggregation processes~\cite{jbj1,jbj2} and protein folding~\cite{chen1,rojas1}\kern0pt. 
  For more details on this work see Ref.~\cite{mbj2}.

  The central quantity is the density of states $g(E)$
  or the microcanonical entropy defined as  $S(E)\equiv \ln\, g(E)$.
  Here, we normalize everything by the number of monomers and use
  \begin{equation}
  s(e)= N^{-1} \ln g(e),
  \end{equation}
  with $e=E/N$.
  In contrast to canonical ($NVT$) statistics, where $T$ is an externally
  fixed control parameter, in the microcanonical ($NVE$) ensemble it is
  \emph{derived} from the entropy, 
  $ T(e)=[\partial s(e)/\partial e]_{N,V}^{-1}$.
  There are cases for finite systems, where $s(e)$ is a convex function in a transition
  regime.
  A consequence is that
  with \emph{increasing} system energy the temperature \emph{decreases}. 
  This is true as long as the surface-to-volume ratio 
  is large enough to suppress a concave increase of $s(e)$.
  In such a case, the energetic separation of the two distinct
  phases is sufficiently large
  to establish a kinetic barrier.
  This regards all first-order phase transitions and
  two-state systems, but also
  transitions, where phase coexistence is completely absent
  in the thermodynamic limit, but not for the finite system.
  The latter is the case here: The adsorption transition
  of flexible polymers to an attractive substrate is known to
  be continuous in the thermodynamic
  limit. However, as we will show here, the adsorption of finite non-grafted polymers
  exhibits signals of a first-order transition
  which vanish in the thermodynamic limit. 

  \begin{figure}[t]
  \begin{center}\includegraphics[width=7.85cm]{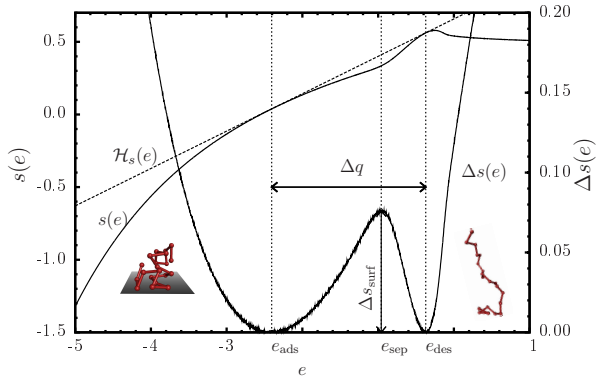}\end{center}
  \caption{\label{fig:s20} Microcanonical entropy $s(e)$ (up to a constant)
  for a 20mer at $\epsilon_s=5$, the Gibbs hull $\mathcal{H}_s(e)$, and the difference
  $\Delta s(e)=\mathcal{H}_s(e)-s(e)$ versus energy per monomer $e$.
  The local maximum of $\Delta s(e)$, called surface entropy
  $\Delta s_{\rm surf}$, defines the energy of phase separation.
  The latent heat $\Delta q$ is defined as the energy being necessary to cross the transition region
  at the transition temperature $T_{\rm ads}$.}
  \end{figure}

  Exemplified for a 20mer and $\epsilon_s=5$, we have plotted
  in Fig.~\ref{fig:s20} the microcanonical entropy $s(e)$.
  It shows the characteristic features of a transition with phase coexistence in
  a small system. For energies right below $e_{\rm ads}$, the system is in the adsorbed phase 
  AE2 (cf.~Fig.~\ref{fig:diagram}), for 
  $e_{\rm ads}< e < e_{\rm des}$, the system is in the transition region, where $s(e)$ is convex. 
  One can construct the Gibbs hull 
  \begin{equation}
  \mathcal{H}_s(e)=s(e_{\rm ads})+e(\partial s/ \partial e)_{e=e_{\rm ads}}
  \end{equation}
  as the tangent that touches $s(e_{\rm ads})$ and $s(e_{\rm des})$,
  whose inverse slope $ T_{\rm ads}=\left({\partial \mathcal{H}_s}/{\partial e}\right)^{-1}$ 
  is the microcanonical \emph{definition} of the adsorption temperature.
  However, the transition rather spans a region of temperatures
  like the fluctuation maxima do in the canonical ensemble.
  Hence, this adsorption temperature definition is not the only one possible.
  A unique transition point only exists in the thermodynamic limit.
  Nevertheless, not only for systems, where the thermodynamic limit is unreachable~\cite{bj4} in principle
  such as for proteins, it is worthwhile to understand the behavior of such a quantity.
  For a further analysis, we also use the surface (or interfacial) entropy, 
  representing the entropic barrier of the transition,
  $ \Delta s_{\rm surf}=\max\{\Delta s(e)=\mathcal{H}_s(e)-s(e)\, |\, e_{\rm ads}\le e\le e_{\rm des}\}$
  and the latent heat, 
  $ \Delta q=e_{\rm des}-e_{\rm ads}$.
  Before we show for the adsorption transition that $ \Delta q$ decreases with $N$,
  we first investigate the origin of the phase separation for finite chains.

  \begin{figure}[t!]
	\begin{center}\includegraphics[width=7.84cm]{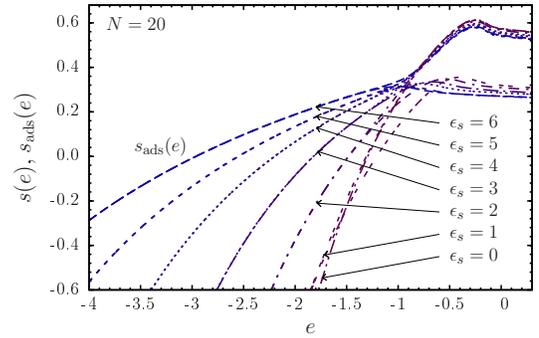}\end{center}
  	\caption{\label{fig:eps} 
  	${s}(e)$ and its fraction for adsorbed conformations $s_{\rm ads}(e)$
  	at various surface attraction strengths $\epsilon_s=0,1,\ldots ,6$ for a 20mer 
  	[for all $\epsilon_s$, the fraction for desorbed structures, $s_{\rm des}(e)$ resembles $s(e)$ for $\epsilon_s=0$].}
  \end{figure}

\textit{Dependence on the Surface Attraction Strength.}
  In Fig.~\ref{fig:eps}, ${s}(e)$ is shown for a 20mer
  and different $\epsilon_s$, where here also $\epsilon_s=6$ has been included. Since the high-energy regime is
  dominated by desorbed conformations, $s(e)$ is hardly affected by 
  $\epsilon_s$ here, while the low-energy tail
  increases significantly with $\epsilon_s$. 
  Thus, one can split the density of states into contributions of 
  desorbed and adsorbed conformations, $g_{\rm des}(e)$
  and $g_{\rm ads}(e)$, respectively,
  such that $g(e)=g_{\rm des}(e)+g_{\rm ads}(e)$ and $s_{\rm des,ads}(e)=N^{-1}\ln g_{\rm des,ads}(e)$.
  We consider the polymer to be adsorbed if 
  $E_{\rm surf}<-0.1\, \epsilon_s\, N$.
  This choice includes all polymers that are reasonably strongly adsorbed in terms of energy and works for all 
  $\epsilon_s$. Moreover, qualitative features do not depend sensitively on the choice
  and this devision is only employed to get a qualitative picture, not to extract $\Delta q$ or $\Delta s_{\rm surf}$.
  Since both, $s_{\rm ads}(e)$ and $s_{\rm des}(e)$, are concave in the whole energy range of
  the transition, the convex entropic monotony can only 
  occur in the region, where adsorbed and desorbed conformations
  have similar entropic weight.
  
  Performing the Gibbs construction as in Fig.~\ref{fig:s20} to extract $\Delta s(e)$, one sees~\cite{mbj2}
  that for $\epsilon_s\gtrapprox 2$ the transition appears to be first-order like ($\Delta q=e_{\rm des}-e_{\rm ads}>0$) for a finite, non-grafted chain.
  For $\epsilon_s\lessapprox 2$, the Gibbs construction is no longer meaningful in absence of a convex regime in $s(e)$, 
  indicating a second-order phase transitions ($\Delta q=0$).
  Referring to the phase diagram in Fig.~\ref{fig:diagram}, the adsorption
  transition thus seems to become first-order-like at that point, where it falls together with the $\Theta$-transition
  ($\epsilon_s\approx 1.8$, $T\approx 1.3$). This is also signaled by the saddle point of the corresponding
  $T^{-1}(e)$ curve.
  For larger $\epsilon_s$, phase coexistence gets apparent 
  between DE and AE2. 
  Here, $\Delta s_{\rm surf}$ and $\Delta q$ increase with $\epsilon_s$
  and trivially diverge for $\epsilon_s\to\infty$.
  Also the first-order-like features of $T^{-1}(e)$
  increase and the adsorption temperatures $T_{\rm ads}$ 
  depend roughly linearly on $\epsilon_s$,
  as was already suggested by the canonical data.

\textit{Chain-length Dependence.}
  Since the adsorption transition   is expected to be of second order
  in the thermodynamic limit\cite{eisenriegler1}, first-order
  signatures found for the finite system between DE and AE2 must disappear for 
  $N\to\infty$. 
  Indeed, our data for $N$ up to $N=150$ 
  support a power-law scaling of the latent heat, $\Delta q \sim N^{-\kappa_q}$,
  with $\kappa_q \approx 0.35-0.40$, which clearly suggests
  $\lim_{N\rightarrow\infty}\Delta q=0$, confirming this expectation.

\textit{Variation of the Box Size.}
  After noticing that there is a considerable influence of the simulation box size on $s(e)$,
  we also investigated this effect. To this end, 
  simulations with fixed $\epsilon_s=5$ and chain length $N=20$ were performed for different
  $L_z=20,30,\ldots, 150$.
  Because the number of adsorbed conformations cannot depend on $L_z$,
  the unknown additive constants to $s(e)$, $s_{\rm ads}(e)$, and $s_{\rm des}(e)$ were 
  chosen such that $s_{\rm ads}(e)$ coincides for all $L_z$. 
  With this choice, $s_{\rm des}(e)$ increases with the logarithm of  $L_z$, like it should be
  the case for the {\em translational} entropy.
  Consequently, both, the surface entropy $\Delta s_{\rm surf}$
  and the latent heat $\Delta q$ increase with $L_z$. 
  Note, that in the case of a grafted polymer, effectively corresponding to a small $L_z$, we did not observe any convex intruder
  in the microcanonical entropy. 

\section{Summary}
  In this work, we have used two approaches to describe the behavior of 
  a single 
  homopolymer near an attractive substrate.

  First, in analyses of canonical expectation values of several energetic 
  and structural quantities and their thermal fluctuations 
  for a chain with $20$ monomers, conformational phases and phase boundaries 
  in the pseudo-phase diagram versus temperature and surface attraction strength were identified.
  Our chosen simulational method was the multicanonical Monte Carlo technique.
  Although the computational expense to accurately explore 
  such a broad parameter range restricted us to rather 
  short chains, 
  for the majority of pseudo-phases, in particular those that are
  assumed to be relevant in the thermodynamic limit, we find a nice qualitative coincidence
  with similar lattice studies.
  Then, we complemented the picture by focusing on the adsorption transition microcanonically.
  For short polymers, the microcanonical entropy revealed that at the adsorption transition 
  adsorbed and desorbed conformations coexist,
  corresponding to a first-order character of this transition for short polymers.
  We have studied how the character of this transition
  depends on surface attraction strength, chain length, and concentration. 

  Altogether, our study has shown the usefulness of a combined approach of the microcanonical and
  canonical ensemble in understanding the conformational behavior of finite systems. \\

\section{Acknowledgements}
This work is partially supported by the DFG (German Science 
Foundation) within the Graduate School BuildMoNa and 
under Grant No.\ JA \mbox{483/24-1/2/3}, the Deutsch-Franz\"osische Hochschule 
(DFH-UFA) under Grant No.~CDFA-02-07, and by the German-Israel Program ``Umbrella''
under Grant Nos.\ SIM6 and HPC\_2. 
Support by supercomputer time grants (Grant Nos.~hlz11, JIFF39, and JIFF43) of the 
Forschungszentrum J{\"u}lich is gratefully acknowledged.
\vspace*{0.5cm}

\end{document}